\documentclass[epj,nopacs]{svjour}
\usepackage{epsfig}
\usepackage{latexsym, amssymb} 
\usepackage{amsmath}
\usepackage{psfrag}
\usepackage{bm}
\usepackage{feynmf}
\usepackage{rotating}
\usepackage{hyperref}
\usepackage{soul}
\usepackage{latexsym}
\usepackage{dcolumn}
\usepackage{amsfonts,amssymb,amsmath}
\usepackage{graphicx,epsfig}
\usepackage{psfrag}
\usepackage{commath}
\usepackage{graphicx}
\usepackage{caption}
\usepackage{subcaption}
\usepackage{cite}
\usepackage[utf8]{inputenc}
\usepackage{float}
\usepackage{authblk}
\usepackage[toc,page]{appendix}

\graphicspath{{./plots/}}

\newcommand{\dd}{\mathrm{d}}
\newcommand{\ee}{\mathrm{e}}

\newcommand{\tw}[1]{\texttt{{#1}}}
\newcommand{\mc}[1]{{\mathcal{#1}}} 

\newcommand{\ie}{\textit{i.e.~}}

\def\beq{\begin{equation}}
\def\efq{\end{equation}}
\def\br{\begin{eqnarray}}
\def\er{\end{eqnarray}}
\def\benu{\begin{enumerate}}
	\def\efnu{\end{enumerate}}

\def\pa{{\partial}}

\def\cR{{\cal R}}

\def\pa{{\partial}}

\def\cR{{\mathcal{R}}}
\def\fnl{{$f_{NL}\,$}}
\def\cP{{\mathcal{P}}}
\def\cC{{\mathcal{C}}}

\def\pl{\rm{Plik}}
\def\CS{\rm{CS}}
\makeatletter
\gdef\@fpheader{}
\makeatother

\begin{document}

\title{\bf Effect of damped oscillations in the inflationary potential}
	\author{Akhil~Antony\inst{1,}\inst{2} \and Shweta Jain \inst{3}}
	\institute{The Institute of Mathematical Sciences, CIT Campus, Chennai 600113, India. \email{akhilantony@imsc.res.in} \and Homi Bhabha National Institute, Training School Complex, Anushakti Nagar, Mumbai 400085, India. \and Department of Physics and Astronomy, University of Kentucky, KY, USA.\email{shweta.jain@uky.edu}} 

\abstract{We investigate the effect of damped oscillations on a nearly flat inflationary potential and the features they produce in the power-spectrum and bi-spectrum.  We compare the model with the Planck data using Plik unbinned and CamSpec clean likelihood and we are able to obtain noticeable improvement in fit compared to the power-law $\Lambda$CDM model. We are able to identify three plausible candidates each for the two likelihoods used. We find that the best-fit to Plik and CamSpec likelihoods match closely to each other. The improvement comes from various possible outliers at the intermediate to small scales. We also compute the bi-spectrum for the best-fits. At all limits, the amplitude of bi-spectrum, \fnl is oscillatory in nature and its peak value is determined by the amplitude and frequency of the oscillations in the potential, as expected. We find that the bi-spectrum consistency relation strictly holds at all scales in all the best-fit candidates.}  	
\maketitle

\tableofcontents

\section{Introduction}
Over the last three decades, tremendous advances in precision cosmology have aided our understanding of the early universe. The Standard Model (SM) has emerged as the most successful model for describing the evolution of the universe owing to the support of numerous precise observations. While SM enjoys widespread acceptance, it does have a few drawbacks, such as horizon problem, flatness problem etc. Among the numerous candidate theories for the early universe, inflation~\cite{Starobinsky:1979ty,Starobinsky:1980te,Guth,Sato,Mukhanov,Linde:1981mu,Linde} has proven to be the best contender to account for these problems. Additionally, inflation has been able to account for the dynamics of primordial fluctuations that seeded the formation of large scale structures today~\cite{Kolb:1990vq,liddle_lyth_2000}. The imprints of these primordial fluctuations can be best identified in the Cosmic Microwave Background Radiation (CMB).

From COBE~\cite{COBE,COBEnormalization} to the PLANCK~\cite{2020,planckcollaboration2020planck,planckcollaboration2019planck} mission, we have made significant progress in our understanding of CMB physics. With the help of CMB data, inflation has emerged as the most promising candidate for describing the near homogeneous and isotropic nature of the Universe over the large scale. Numerous inflationary models exist that appear to be consistent with the CMB observations, in which a near flat potential generates a nearly scale invariant spectrum of scalar perturbations. While these models appear to be consistent with the observational data, it has been noted that adding a few features to these flat power spectra may result in a better fit to the data~\cite{Starobinsky:Kink,Ivanov:1994pa,Adams:Step0,Covi:Step1,Allahverdi:PI,Ashoorioon:Step2,Joy:2007na,Joy:2008qd,Jain:PI,Pahud:Osc2,Hazra:Step3,McAllister:Osc0,Flauger:Osc1,Miranda:Step4,Aich:Osc3,Peiris:Osc4,Benetti:Step5,Meerburg:Osc5,Easther:Osc6,Bousso:Step8,GallegoCadavid:Step6,Chluba:Step7,Motohashi:Osc7,Miranda:Osc8}. These additional features are found to fit a few consistent outliers that are being observed in the data for decades~\cite{Hannestad_2001,Tegmark_2002,Shafieloo_2004,Hazra_2013,Nicholson_2009,Nicholson_2010,SL_Bridle,Shafieloo_2007,Shafieloo_2008,Hazra:recon13broad,Hazra:reconP13,Hunt:2015iua,Obied:2018qdr,Braglia:2020fms,Braglia:2021ckn,Hazra:2021eqk,Braglia:2021rej,Braglia:2021sun}. The persistent existence of these outliers could be owing to some unknown features of inflationary dynamics, which could expand our understanding of the early universe. As a result, it is critical to investigate about these features in the power spectrum. We provide a single field canonical inflationary potential in this study that simulates these extra features using damped sinusoidal oscillations. A noteworthy aspect of this feature is that it could generate both sharp and resonant featured oscillations, as well as their combinations~\cite{Chen:2016zuu}. Primordial standard clock~\cite{Chen:2014cwa} is one model that can generate a combination of sharp and resonant spectral features, but it is a two field inflationary model. It is shown in \cite{Antony:2022ert} that these features could account for the additional smoothing in the CMB temperature spectrum, thus resolving the $A_{lens}$ anomaly. It increases $H_0$ while decreasing $S_8$. Additionally, the model fits the One spectrum~\cite{Hazra:2022rdl}, which resolves many tensions and anomalies in the Planck data. In this study, we examine alternative best fit candidates for this feature, in addition to the one mentioned in \cite{Antony:2022ert}, by analysing the entire $k$-space. Using Planck CMB data, we identify these features in the primordial power spectrum using this range of spectra. Later, we will demonstrate that these features could account for some of the in  outliers CMB data that are not captured by the power-law form of the primordial power-spectrum. We examine a few possible candidates, sharp featured and resonant featured, that improve the fit to CMB data noticeably.

This paper is structured as follows. In \autoref{Meth}, we introduce the potential form and discuss the methodology of work as well as the various datasets used for it. The results and discussion of various best-fit candidates are provided in \autoref{Res}. Finally, we discuss the conclusions and inferences drawn from the work in \autoref{conc}. Appendices contain the analytical calculations used in the work. We used natural units throughout the paper, $\hbar= c=M_{pl}=1$. In this paper, we use the metric signature (-,+,+,+). For formalising the equations, we used three sets of coordinates for time, namely cosmic time $(t)$, conformal time $(\eta)$, and the number of e-folds $(\mc{N})$. We work in an expanding homogeneous and isotropic universe whose metric is given by the Friedmann Lemaitre Robertson Walker (FLRW) metric having perturbations up to linear order. An over-dot and an over-prime denote differentiation in terms of $t$ and $\eta$, respectively. Differentiation with respect to $\mc{N}$ is given by a subscript $\mc{N}$, \ie $f_{\mc{N}} = \dd f/\dd \mc{N}$.

\section{Model and data} \label{Meth}
In this paper, we investigate the dynamics of a single canonical scalar field in an inflationary potential. Our potential is divided into two parts: a baseline with a slow roll regime where we could begin the inflation, and a small feature in the form of damped cosine oscillation. The nearly scale invariant (slightly red titled) power-spectrum is obtained from the baseline part of the potential. We add features in this baseline potential to capture extra possible signals in CMB data. The feature we propose could be added to any baseline potential that could produce a near scale invariant power spectrum. This was verified using a couple of different base potentials. One can also directly add features to the power-spectrum and estimate parameters, but in such cases, it may be difficult to obtain an expression for potential back from the spectra. Therefore, we work with features in the potential itself.

\begin{figure*}[ht]
\includegraphics[trim={0 0 0 0},width=\textwidth]{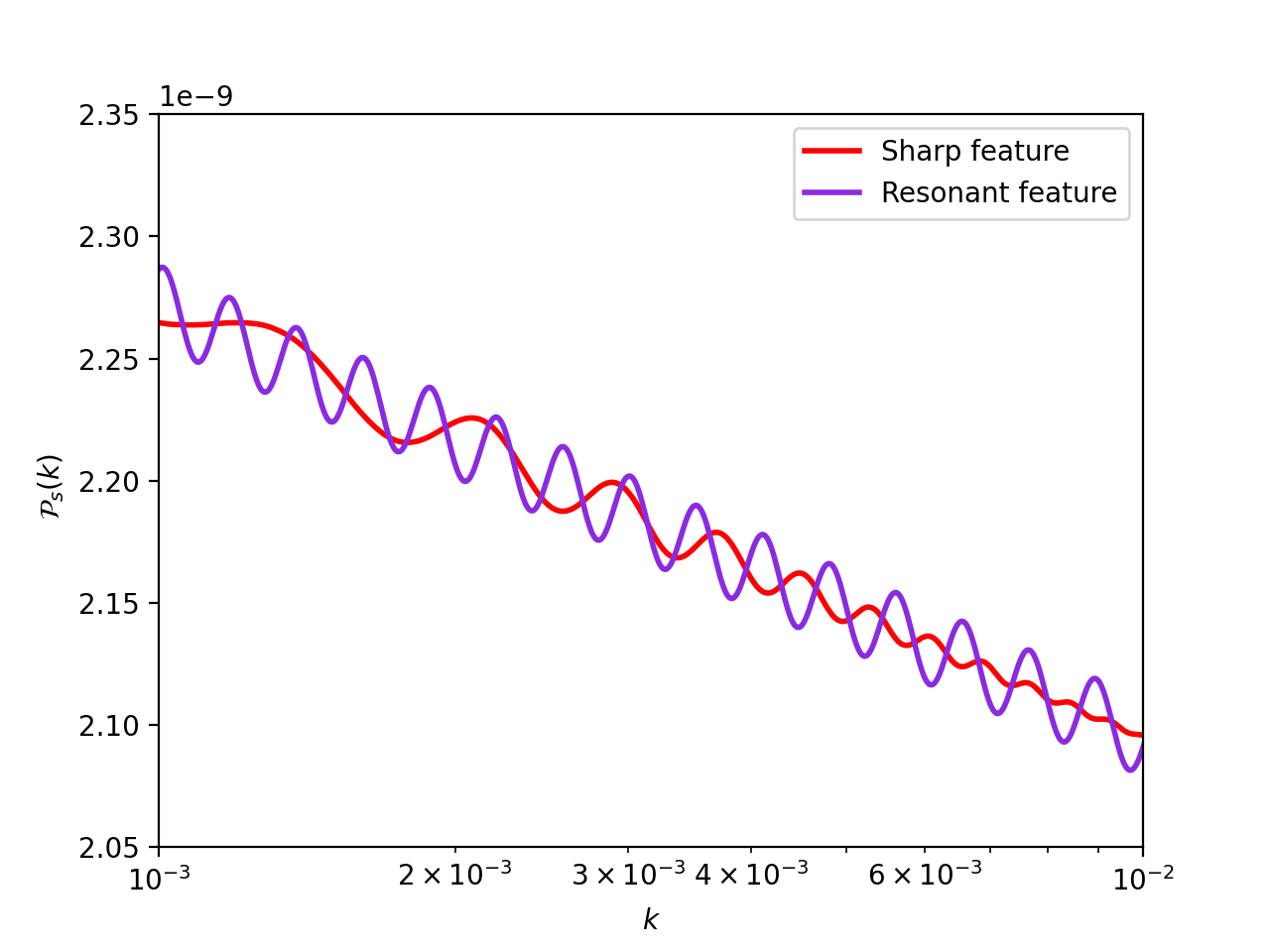}
\caption{ This figure shows  two types of features that can be generated by our model in different limits. Here, the red curve is the sharp featured oscillation  whose peaks are separated linearly in k and purple curve corresponds to the resonant features whose peaks are separated uniformly in $\log k$. Note that the x-axis is in $\log$ scale here. }
\label{sharp_reson}
\end{figure*}

We consider the potential to be of the following form,
\beq{
V(\phi) = \gamma^2 \phi \cosh{(\beta_1 \phi)} + \frac{\alpha\cos{(\omega_\phi \phi)}}{[\beta_2(\phi-\phi_0)]^2 +C}. \label{pot}
}
\efq
Here $\beta_1$ and $\beta_2$ are fixed and $C$ is an arbitrary constant added to avoid the divergence at $\phi=\phi_0$. It can be any value other than $0$. The power-spectrum tilt is controlled by $\beta_1$. We fix $\beta_1$ by performing a parameter estimation with only baseline potential parameters, \ie by varying only two parameters $\gamma$ and $\beta_1$. We allow $\beta_1$ to vary between values that result in an approximate tilt of $0.94-0.98$. This analysis done using {Plik bin1}. The best-fitting value for $\beta_1$ results in a tilt of $\sim0.96$. $\beta_2$ regulates the amount of damping applied to the oscillations, ensuring that features remain localised. We can see from \autoref{sharp_reson} that using the above potential, we can produce both sharp ($\omega_\phi\rightarrow 0$) and resonant features ($\beta_2\rightarrow 0$)in the power-spectrum.

\begin{figure*}[ht]
\includegraphics[trim={2cm 0 0 0},width=\textwidth]{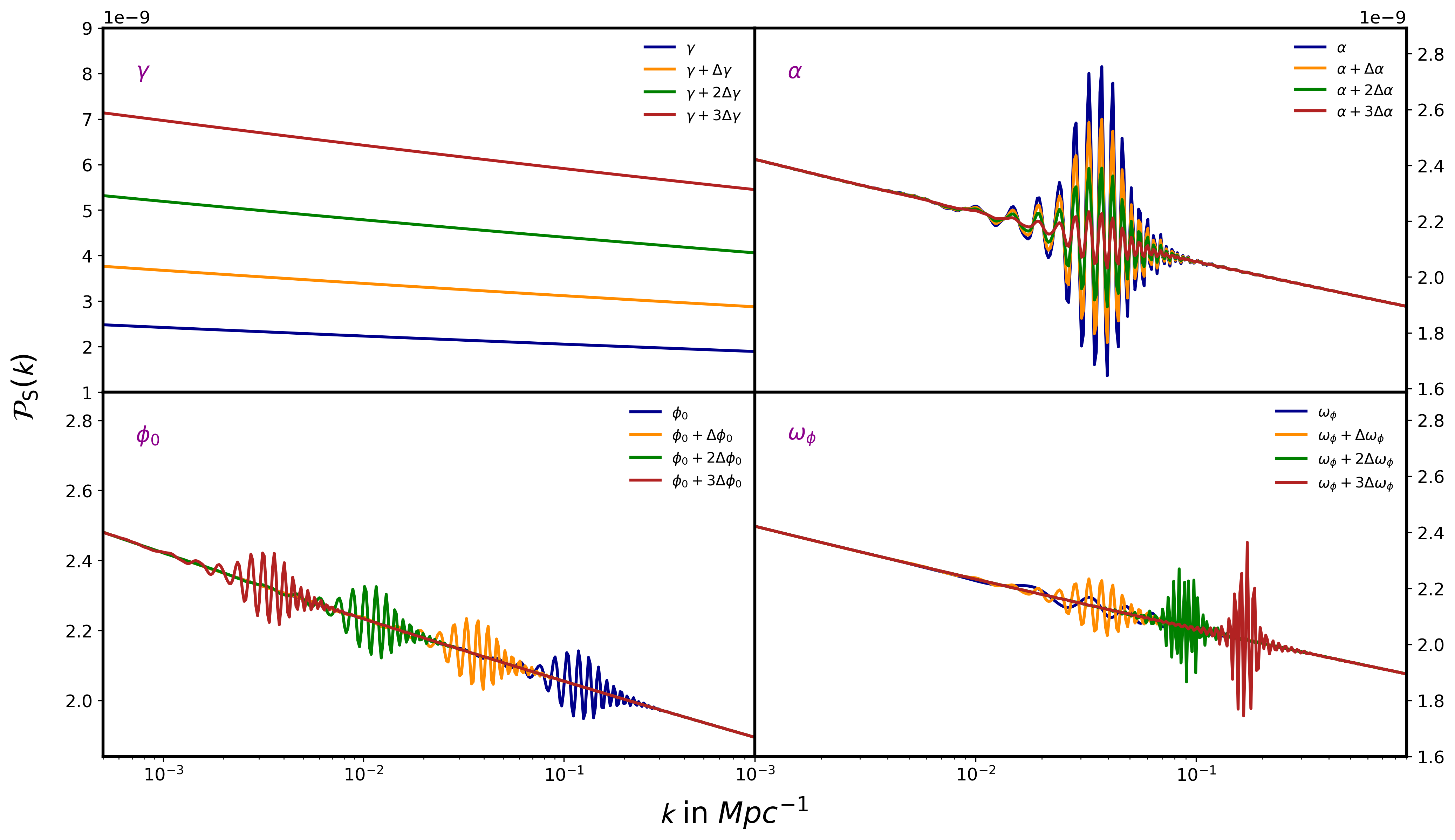}
\caption{Variation of the power spectrum in response to changes in potential parameters. Here, we varied one parameter by a specific step size while holding the other three constant. In the upper right corner of each panel, the varying parameter is specified.}
\label{PS-vary}
\end{figure*}

We vary four parameters in the full potential function, one in the baseline part and three in the feature part. The effect of each parameter on the power-spectrum could be understood from \autoref{PS-vary}.
Potential, as described in \autoref{pot}, contains sinusoidal oscillations which damps as we move both sides from $\phi_0$. Assuming a slow roll dynamics during the initial phase of inflation, given an initial value for the field($\phi_i$), one can obtain other initial conditions required to calculate the background dynamics. Here we are working with number of e-folds ($\mc{N}$), defined as  $a=a_i\ee^\mc{N}$ where $a$ is the scale factor. Initially, the potential term dominates the kinetic term therefore one can approximate Hubble parameter as, $3H^2 \approx V(\phi)$. Therefore, $(\phi_\mc{N})_i$   for a given $\phi_i$ evaluates to be,
\begin{subequations}
\begin{eqnarray}
    H_i = \sqrt{V(\phi_i)/3}, \\
    {\phi_\mc{N}}_i = -\frac{V_\phi(\phi_i)}{3H_i}.
    \end{eqnarray}
\end{subequations}
Once we have the initial conditions we solve for the background equations and get the form of $\phi$, $\phi_\mc{N}$, and $H$ as a function of e-folds($\mc{N}$). We do all these calculations using the FLRW metric. With aforementioned quantities, we can get the slow roll parameter, $\epsilon(\mc{N})$, and calculate the end of inflation($\epsilon(\mc{N}_{end})=1$). To find initial value of scale parameter, $a_i$, we impose that the mode $k=0.05 ~{\rm Mpc^{-1}}$ crosses the horizon at $\mc{N}= \mc{N}_{end}-50$ \cite{Hamann:2007pa}. Once the background is evaluated completely, we add the perturbations to the fields and solve for the curvature perturbation as discussed in \autoref{th-inf}

The field continues to be in the slow roll phase till it feels the effect of  oscillations. Once the effect of oscillations take over, the field accelerates into an intermediate fast-roll phase which is  responsible for the features in power-spectrum. Inflation continues as the field rolls further down the potential till $\epsilon$ becomes one. Assuming Bunch-Davies initial condition \cite{Bunch-Davies}, one solves for the curvature perturbation ($\cR$),
\begin{equation}
    \cR_k'' + 2\bigg(\frac{z'}{z}\bigg)\cR_k' + k^2\cR_k = 0,
\end{equation}
and get the power-spectrum from 
\begin{equation}
    \mc{P}_s(k) = \frac{k^3}{2\pi^2}\abs{\cR_k}^2. 
\end{equation} 
One can also use the Mukhanov-Sasaki equation (\autoref{MSE-N2_1}) to get the power-spectrum.

We calculate the primordial power-spectrum numerically with the help of publicly available code $\tw{BINGO}$~\cite{Hazra_2013_BINGO}. $\tw{BINGO}$ solves for $\cR$ for each $k$ to get $\cP_s$ as a function of $k$. Technically, one needs to integrate curvature perturbation throughout the inflationary epoch. But it can be safely approximated to a region from $\mc{N}_i$ deep inside Hubble radius($ k \gg aH$) and to an $\mc{N}_e$ well outside the Hubble radius($k \ll aH$). This could be calculated using the following conditions:
\begin{subequations}
\begin{eqnarray}
    k = &&{C}_{IC} a(\mc{N}_i)H(\mc{N}_i), \\
    k = &&{C}_{SHS}a(\mc{N}_e)H(\mc{N}_e). 
\end{eqnarray}
\end{subequations}
For each mode, we calculate the $\mc{N}_i$ and $\mc{N}_e$ using appropriate choice of ${C}_{IC}$ and ${C}_{SHS}$ values \cite{Salopek:1988qh}. We fix $C_{SHS}$  but we do vary $C_{IC}$ depending on the oscillations present at a given scale, \ie for modes near high frequency oscillations in power-spectrum, we use a larger value for $C_{IC}$. Typically, it's value is 200.

We incorporate $\tw{BINGO}$ into $\tw{CAMB}$ \cite{Lewis_2000,CAMB} and calculate the angular power-spectrum using the Boltzmann equations. We run Markov Chain Monte Carlo (MCMC) using \tw{CosmoMC} \cite{Lewis_2002,COSMOMC} and identify regions in parameter space which gives an improvement in fit. Starting from the aforementioned regions, we then run BOBYQA \cite{BOBYQA} to obtain the best-fit values for the parameters.

We performed our analysis using the Planck mission's most recent CMB temperature and polarisation anisotropy data. Planck was able to map the CMB sky over a wide range of multipoles ($\ell=2-2500$) on both small ($\ell\ge 30$) and large ($\ell=2-29)$ scales. We use two sets of likelihood for high-$\ell$ in our analysis: {Plik-bin1-TTTEEE} \cite{Planck:2019nip,2020} and {CamSpec-v12-5-HM-cln-TTTEEE} \cite{efstathiou2020detailed}.  We use {commander-dx12-v3-2-29} for low-$\ell$ TT and {simall-100x143-offlike5-EE-Aplanck-B} for low-$\ell$ EE. {Plik-bin1-TTTEEE} represents the completely unbinned TTTEEE likelihood, and CamSpec  represents the newly cleaned CamSpec. {CamSpec-v12-5-HM-cln-TTTEEE} employs a sophisticated data analysis pipeline to generate an improved CamSpec likelihood and also an increased sky fraction for temperature and polarisation. In our study, we vary the nuisance parameters in addition to the background parameters, and we include the priors involved, as indicated in the Planck 2018 and CamSpec likelihood papers \cite{2020,efstathiou2020detailed}. The prior used for the potential parameters is given in \autoref{priors}. We perform the analysis across multiple parameter templates and narrow it down to three best-fit candidates for each likelihood. The following section will discuss these three candidates in greater detail.
   \begin{table*}
   \centering
	$$ 
	\begin{tabular}{|p{0.2\textwidth}|p{0.15\textwidth}|}
	\hline
	Parameters & Prior  \\
	\hline

	$\gamma(\times10^{4})$&0.14 - 0.20\\
	$\alpha(\times10^{11})$&-2.0 - 2.0\\
	$\phi_0$& 12.9 - 13.9 \\
	$\omega_{\phi}$& 0 - 529 \\
	\hline
	\end{tabular}
	$$ 
	\caption[]{The table contains the prior range we used for the inflationary potential parameters}
	\label{priors}
\end{table*}

\begin{table*}
   \centering
  
	$$ 
	\begin{tabular}{|p{0.6\textwidth}|p{0.3\textwidth}|}
	\hline
	\texttt{CamSpec-v12-5-HM-cln-TTTEEE}\newline 
	Low-T (\tw{commander-dx12-v3-2-29})\newline Low-E (\tw{simall-100x143-offlike5-EE-Aplanck-B})& \vspace{0.2cm} CamSpec clean(\CS)  \\
	\hline
	\tw{Plik-bin1-TTTEEE} \newline
	{Low-T} (\tw{commander-dx12-v3-2-29}) \newline
	{Low-E} (\tw{simall-100x143-offlike5-EE-Aplanck-B})& \vspace{0.2cm}Plik-bin1(\pl)\\
	\hline
	\end{tabular}
	$$ 
	\caption[]{Short hands used for the combination of Planck likelihoods used for the analysis. We have used same low $\ell$ (lowT+lowE) likelihoods for both the sets.}
	\label{short}
\end{table*}
\section{Results} \label{Res}
The following is the nomenclature used to identify the candidates: The likelihood against which it is tested followed by the candidate number. For example, \pl-1 specifies that it's the first candidate analysed against \pl\,likelihood.

\subsection{Best-fit Candidates}
\begin{figure*}
\includegraphics[trim={0 0 0 0},clip, width=\textwidth]{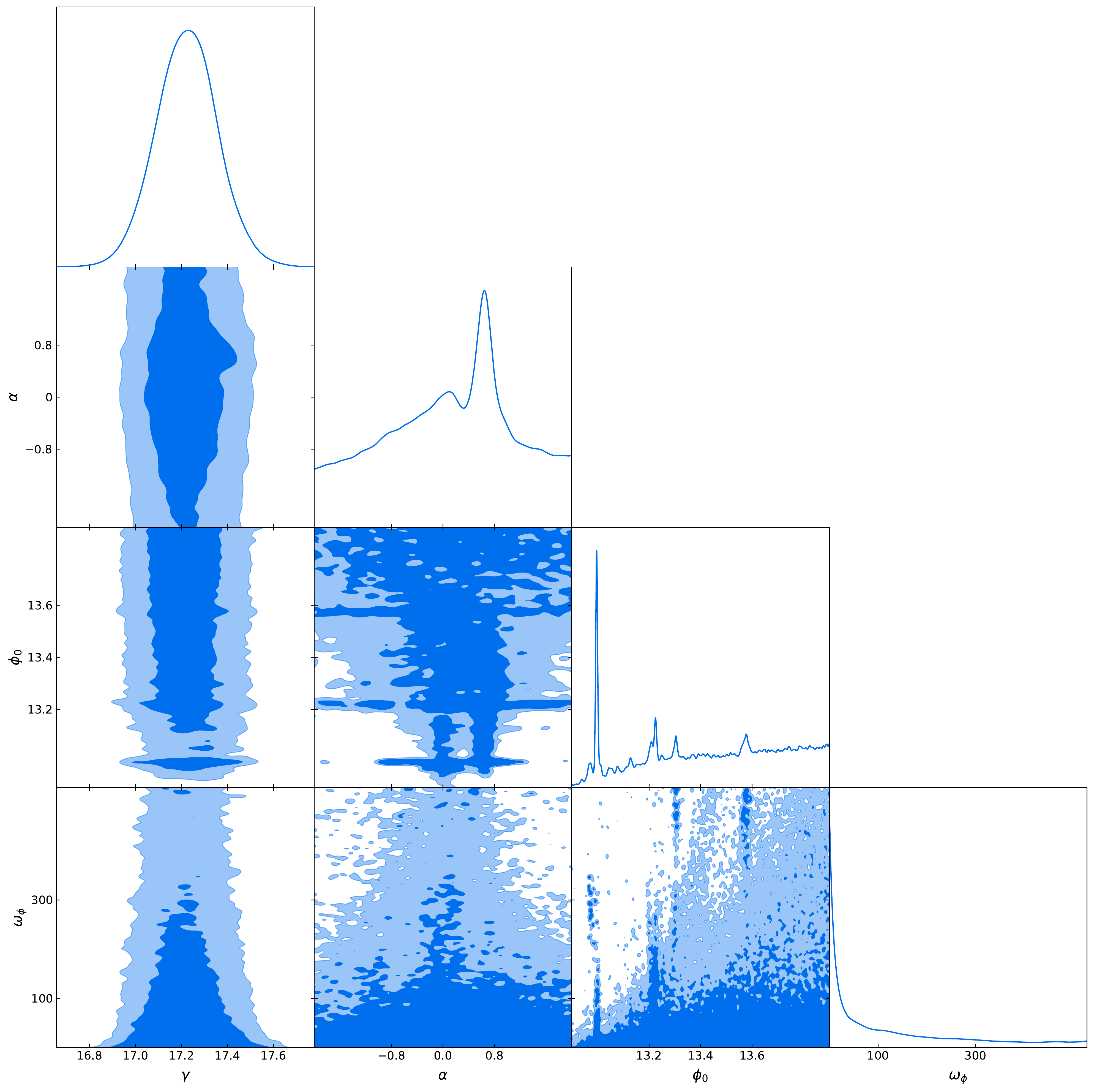}
\caption{ The figure contains the triangle plot for the feature model parameters corresponding to Plik bin1. Here we can see multiple region in parameter space which could lead improved fit to data using different combination of these parameters.}
\label{mcmc}
\end{figure*}

We explored the parameter space using \rm{MCMC} algorithm and were able to identify multiple regions that could improve the fit to data (\autoref{mcmc}). We perform the analysis using Plik-bin1.  To identify these regions, we study various points that lie within $\Delta\chi^2=1$ region from the global minima of $\chi^2$ value of the \rm{MCMC} run using \rm{BOBYQA}. While performing the \rm{BOBYQA} analysis we make use of both Plik-bin1 and CamSpec clean likelihoods. Using \rm{BOBYQA} analysis, we are able to identify three candidates that gave improvement in fit for each likelihood. Those are \pl-1, \pl-2, \pl-3 (\pl\, candidates) and \CS-1, \CS-2, \CS-3 (\CS\, candidates).  We also calculated the bayesian evidence using MCEvidence\cite{2017arXiv170403472H} python package.  The evidence only provided a 0.3-factor weak support for the baseline model. But according to Jeffrey's scale, this is an inconclusive evidence.

\begin{figure*}
\includegraphics[trim={0 0 0 0},clip, width=\textwidth]{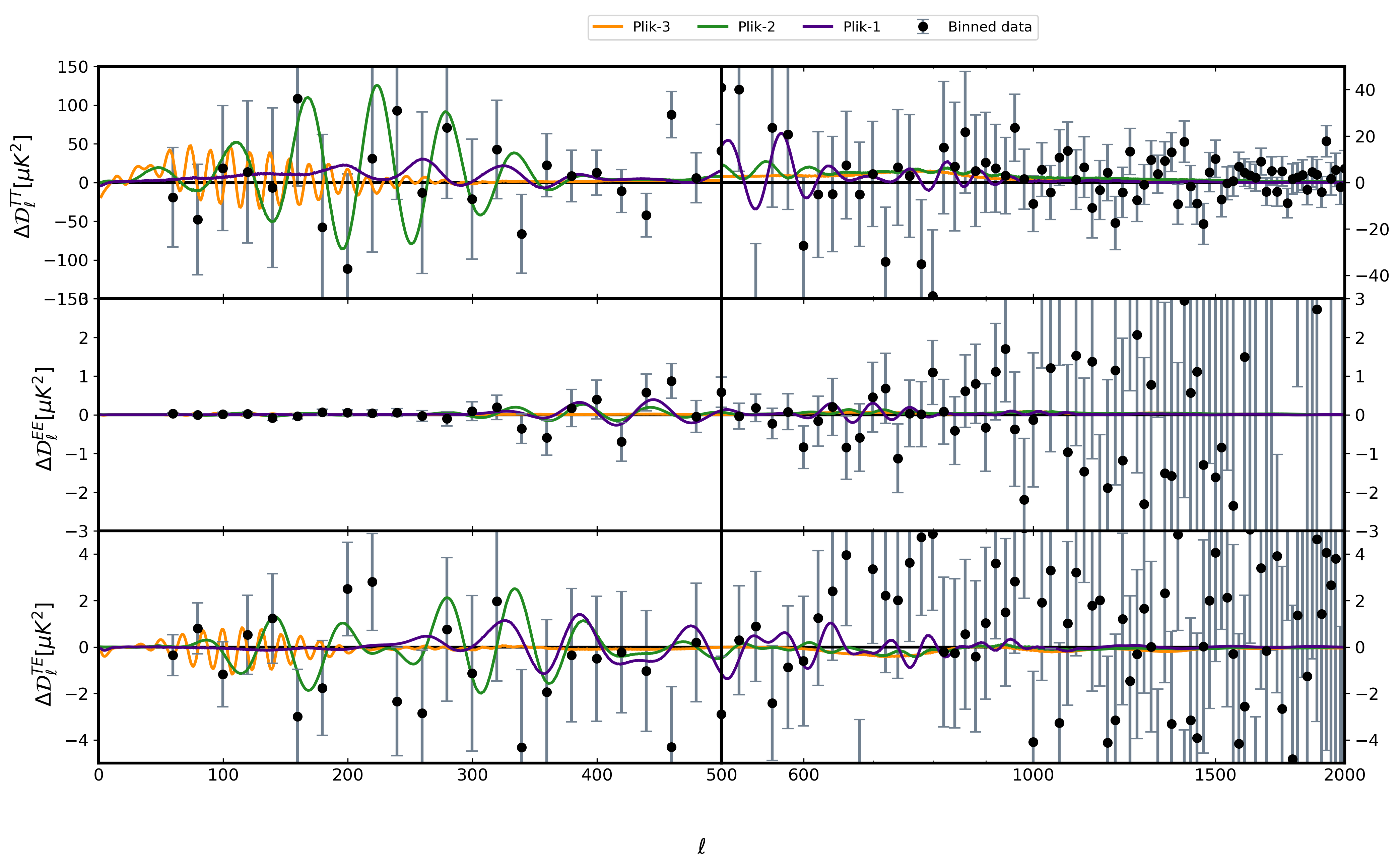}
\caption{Residual plots corresponding best-fit candidates to Plik-bin1 TTTEEE+lowT+lowE likelihood. Here residual is calculated from the power-law $\Lambda$CDM model. Zero line corresponds to the power-law $\Lambda$CDM model and data points are from the 2018 Planck Plik data residual to the power law best fit.}
\label{plik-DL}
\end{figure*}

The best-fit values for the \pl \,runs are presented in \autoref{bf-plik}. We investigated three candidates for the \pl-runs based on the $\chi^2$ improvement we obtained from the \rm{BOBYQA} runs. We are comparing it to the $\chi^2$ obtained from the power-law form of the primordial spectrum (referred to as the power-law model from now on), which has a $\chi^2=24548.5$ value. We can see from \autoref{bf-plik} that we get 10, 8.5, and 6 improvement for the candidates \pl-1, \pl-2, and \pl-3, respectively. The residual plot from \pl-runs with respect to the power-law model is shown in \autoref{plik-DL}. The power-law model is represented by the zero line, and the coloured lines are candidates. We can observe slight power suppression in the case of \pl-3. Here the improvement comes mainly from the high-$\ell$, ~4.3 while the low $\ell$ gives around 1.45 improvement. \pl-3 also gave 2.  The TT residual plot is able to capture the outliers in the range $ \ell = 1000-1500$, while the EE and TE residuals could capture comparatively lower multipoles ranging from 140 to 600. The exact outlying multipole values captured by the \pl-candidates are given by the \autoref{mpv}.
   \begin{table*}
   \centering
	$$ 
	\begin{tabular}{|p{0.2\textwidth}|p{0.15\textwidth}|p{0.15\textwidth}|p{0.15\textwidth}|}
	\hline
	Parameters & \pl-1 & \pl-2 & \pl-3  \\
	\hline
	$\Omega_bh^2$&0.0223 &0.0223&0.0222  \\
	$\Omega_ch^2$&0.1200 &0.1204& 0.1207 \\
	$100\theta_{MC}$& 1.0409&1.0408&1.0409\\
	$\tau$&0.0560&0.0585& 0.0510\\
	\hline
	$\gamma(\times10^{4})$&0.1724&0.1729 &0.1716\\
	$\alpha(\times10^{11})$&-0.1271 &0.6378 &-1.223\\
	$\phi_0$& 12.97&12.99&13.20 \\
	$\omega_{\phi}$&286.96 &3.13&132.02 \\
	\hline
	$-2\log(\mathcal{L})$&24538.47& 24539.98 &24542.57\\
	\hline
	\end{tabular}
	$$ 
	\caption[]{Best-fit parameters and likelihood obtained for the candidates of the Plik-bin1 TTTEEE+lowT+lowE likelihood. First four are the  $\Lambda$CDM background parameters and the next four are the inflationary potential parameters. Final row gives the $\chi^2$ values obtained by the candidates. Improvement in fit obtained are 10, 8.5, and 6 respectively for the candidates \pl-1, \pl-2, and \pl-3.}
	\label{bf-plik}
\end{table*}

\begin{figure*}
\centering
\includegraphics[trim={0 0 0 0},clip,,width=\textwidth]{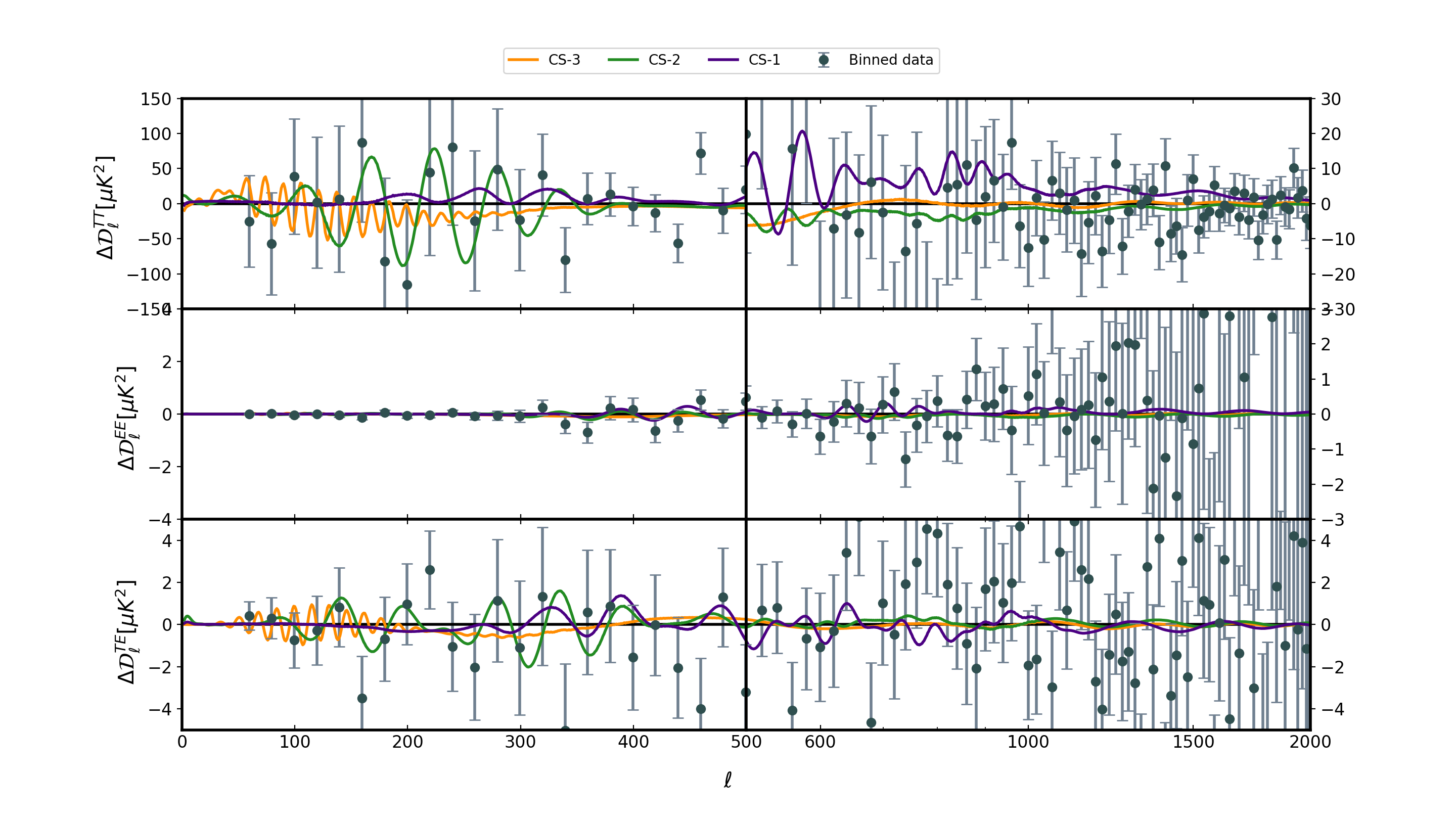}
\caption{Residual plots corresponding best-fit candidates to CamSpec clean TTTEEE+lowT+lowE likelihood. Here residual is calculated from the power-law $\Lambda$CDM model. Zero line corresponds to the power-law $\Lambda$CDM model and data points are from the 2018 Planck CamSpec data residual to the power law best fit.}
\label{CS-DL}
\end{figure*}

The best-fit values for the CS runs are given in \autoref{bf-CS}. The $\chi^2$ value for power-law primordial power-spectrum is $10211.3$, \ie the improvement for candidates \CS-1, \CS-2, \CS-3 are $5.0, 3.8$ and $3.7$ respectively. \autoref{CS-DL} contains the residual plots for CS runs with respective to the power-law model. Similar to \pl-candidates,  residual plot of CS candidates for TT also captures the outliers in large $\ell$ values and EE captures for $\ell\sim 400$. The complete list  $\ell$ values of the outliers captured by the CS candidates are given in \autoref{mpv}.

\begin{table*}
	$$ 
	\begin{tabular}{|p{0.2\textwidth}|p{0.15\textwidth}|p{0.15\textwidth}|p{0.15\textwidth}|}
	\hline
	Parameters & \CS-1 & \CS-2 & \CS-3  \\
	\hline
	$\Omega_bh^2$&0.0222 &0.0222& 0.0221  \\
	$\Omega_ch^2$&0.1205 &0.1200&0.1208\\
	$100\theta_{MC}$&1.0406&1.0410&1.0409\\
	$\tau$&0.0545&0.0602&0.0524\\\hline
	$\gamma(\times10^{4})$&0.1723&0.1728 &0.1717\\
	$\alpha(\times10^{11})$&-0.1021&0.6394& -1.224\\
	$\phi_0$&12.97&12.99 &13.21\\
	$\omega_{\phi}$&294.8 &4.05& 132.25\\\hline
	$-2\log(\mathcal{L})$&10206.29 & 10207.51 &10207.55\\
	\hline
	\end{tabular}
	$$ 
	\caption[]{Best-fit parameters and likelihoods obtained for the candidates of the CamSpec clean TTTEEE+lowT+lowE likelihood. First four are the  $\Lambda$CDM background parameters and the next four are the inflationary potential parameters. Final row gives the $\chi^2$ values obtained by the candidates. Improvement in fit obtained are 5, 3.8, and 3.7 respectively for the candidates CS-1, CS-2 and CS-3.}
	\label{bf-CS}
\end{table*}

\begin{table*}
\centering
\begin{tabular}{|c|c|c|c|}
\hline
RUNS   & TT                   & EE            & TE                   \\ \hline
Plik-1 & 500, 520      & 60, 420, 440 & 500                  \\ 
Plik-2 & 500,1080,1340,1580   & 60, 140, 440 & 160, 200,240,260,500 \\ 
Plik-3 & 1080, 1340           & 60             & 240                  \\ \hline
CS-1   & 520, 960, 1240, 1400 & 420 & 640                    \\ 
CS-2   &1000, 1140,1420,1520                & 420           & -                \\ 
CS-3   & -                & 340           & -                 \\ \hline
\end{tabular}
\caption{Outliers addressed by the Plik-bin1 (first three) and CamSpec clean (last three) candidates from the residual plots of TT, EE and TE correlation in \autoref{plik-DL} and \autoref{CS-DL}. }
\label{mpv}
\end{table*}
\subsection{Scalar power-spectrum} \label{PS}
\begin{figure*}
\centering
\includegraphics[trim={0 0 0 0},width=\textwidth]{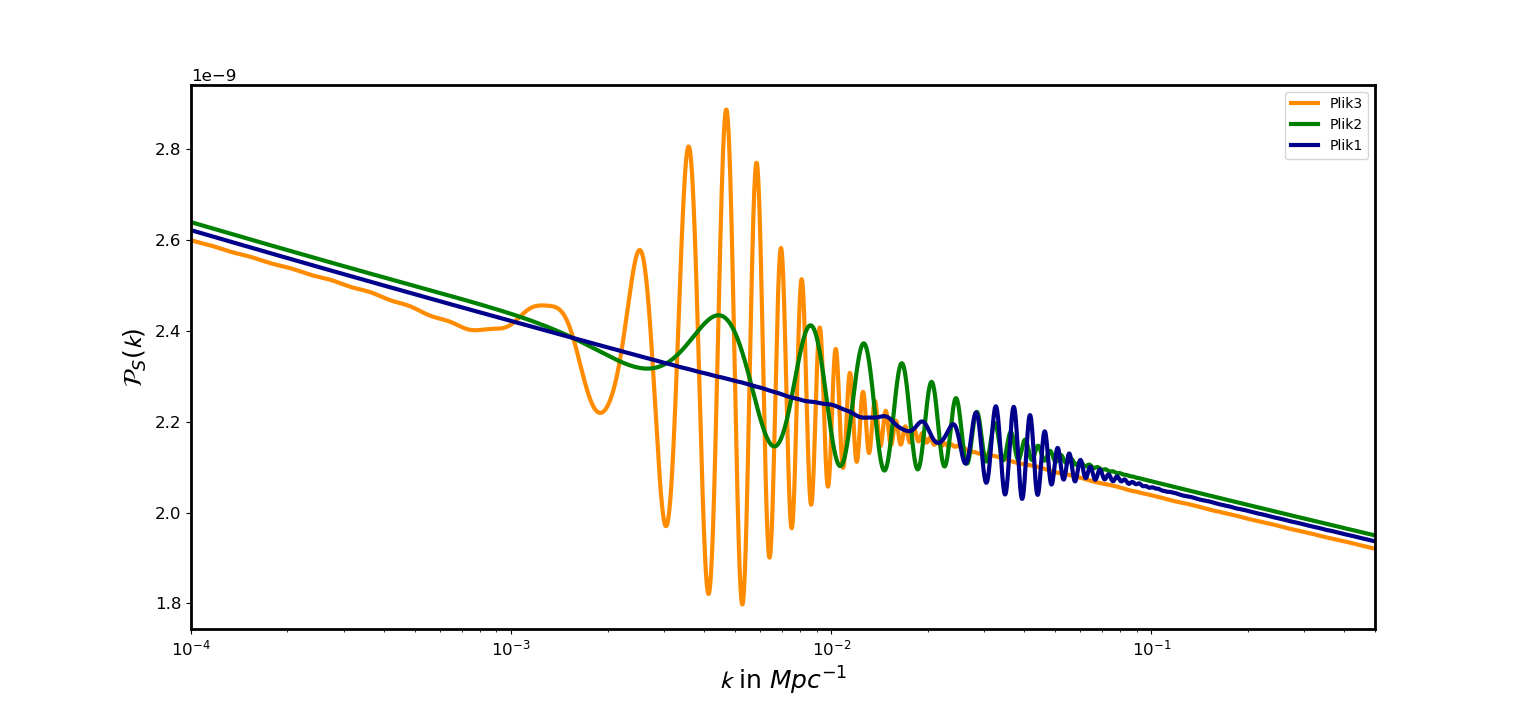}
\caption{power-spectrum for the best-fit candidates to the Plik-bin1 likelihood \pl-1, \pl-2 and \pl-3.}
\label{PS-plik}
\end{figure*}

\begin{figure*}
\centering
\includegraphics[trim={0 0 0 0},width=\textwidth]{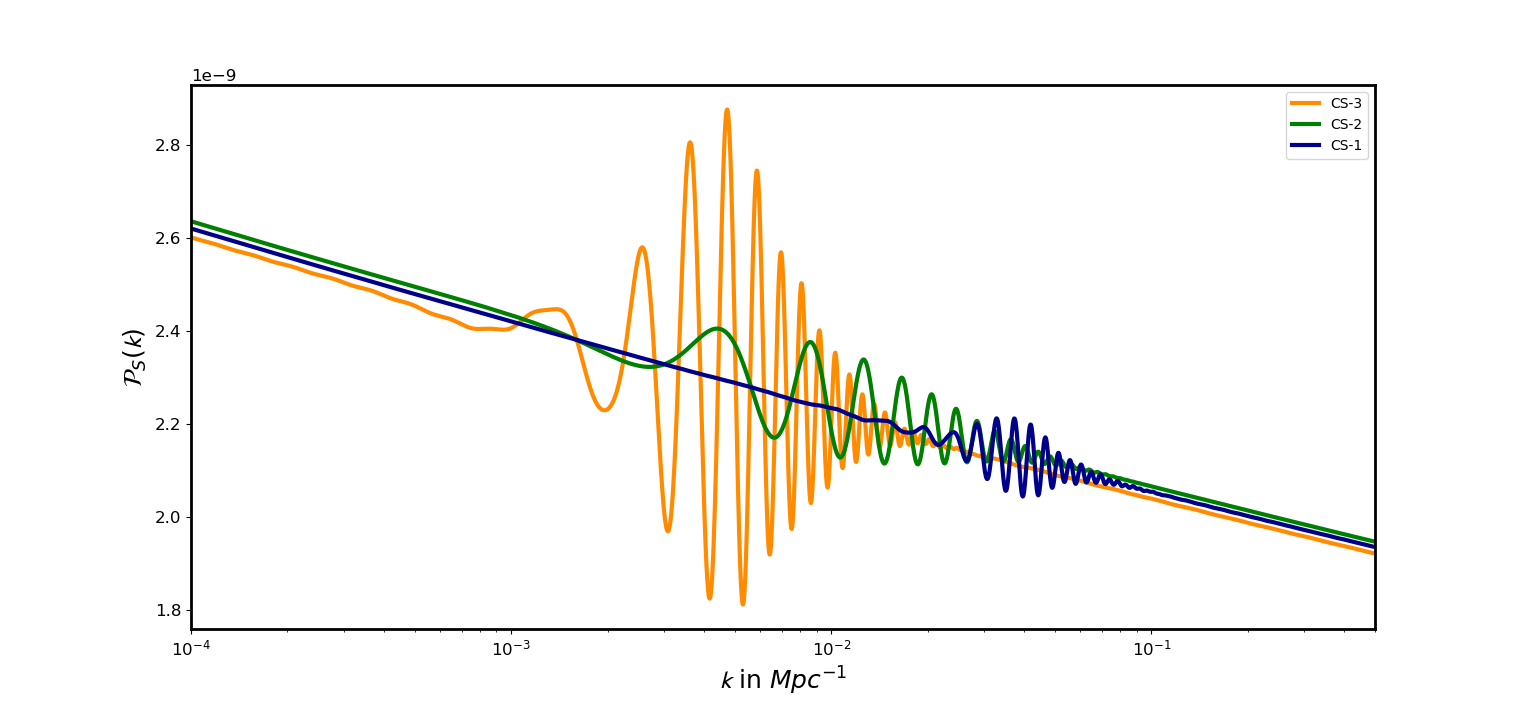}
\caption{power-spectrum for the best-fit candidates to the CamSpec clean likelihood CS-1, CS-2 and CS-3. }
\label{PS-CS}
\end{figure*}

\begin{figure*}
\centering
\includegraphics[trim={0 0 0 0},width=\textwidth]{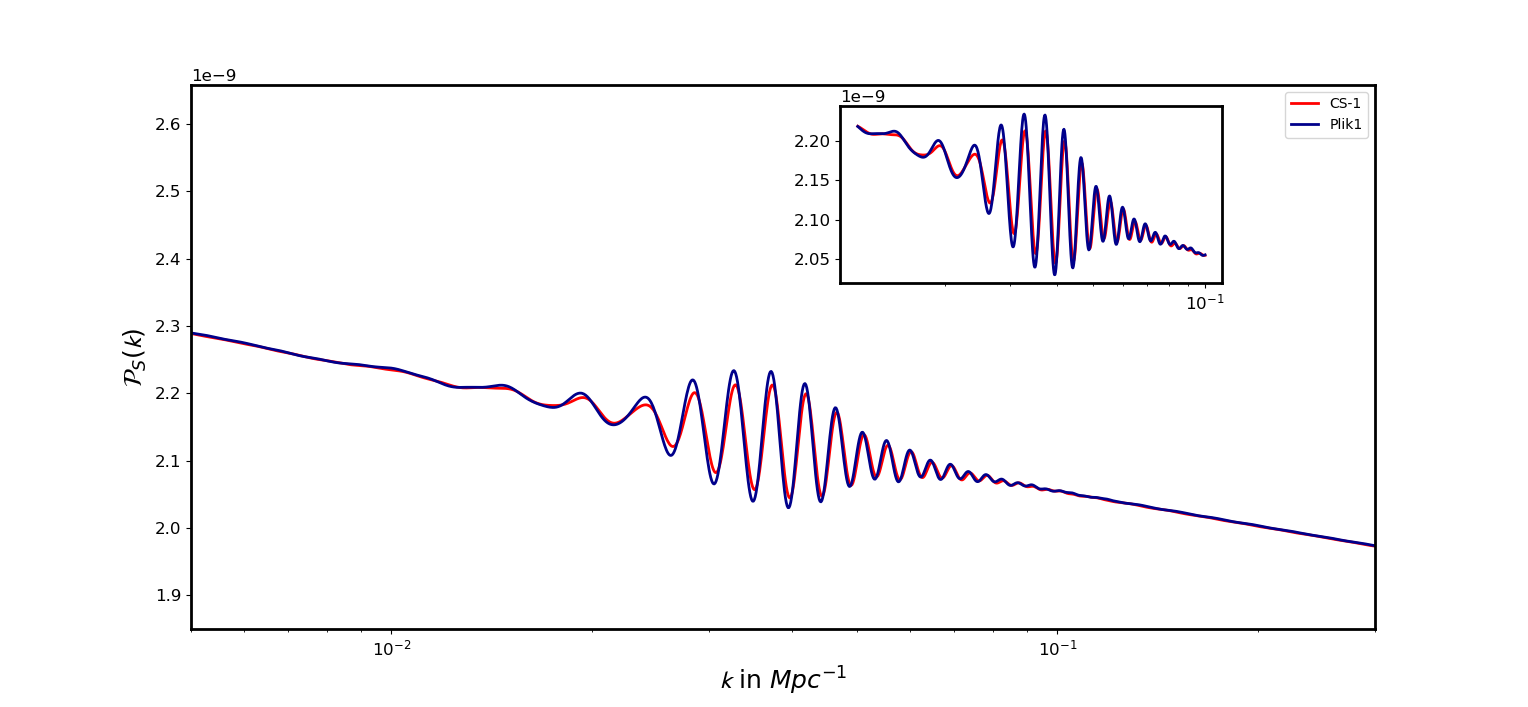}
\caption{power-spectrum for the global best-fit of both CamSpec clean (CS-1) and Plik-bin1 (\pl-1) likelihood. }
\label{global-bf}
\end{figure*}

We present here the local and global best-fits to the data. We saw an improvement in $\chi^2$ at three different points in the parameter space. One for low frequency, one for intermediate frequency, and one for high frequency oscillations. CS-1 and \pl-1, which have a high frequency, are the global best-fit to the data for both CamSpec clean and Plik-bin1 likelihoods. They are both located in very close proximity in the parameter space. \pl-1, the best-fitting Plik candidate, has features at a smaller scale, $k \sim 4\times10^{-2} ~{\rm Mpc^{-1}}$, whereas \pl-3 has the features at slightly larger scales that end near $k \sim 10^{-2} ~{\rm Mpc^{-1}}$. \pl-2 have features ranging from $k \sim 10^{-3} ~{\rm Mpc^{-1}}$ to $k \sim 5\times10^{-2} ~{\rm Mpc^{-1}}$. We saw a similar pattern with the CamSpec clean candidates, which are very close to the Plik candidates in the parameter space and in the same order. The power-spectrum for both sets of candidates has been plotted in \autoref{PS-plik} and \autoref{PS-CS}. In \autoref{global-bf}, a comparison of the global best-fit for two sets of candidates is given. Both are found in the same location, and the amplitude and frequency of the oscillations are comparable.

\subsection{Scalar bi-spectrum}
\begin{figure*}
\centering
\includegraphics[trim={0 0 0 0},width=0.9\textwidth]{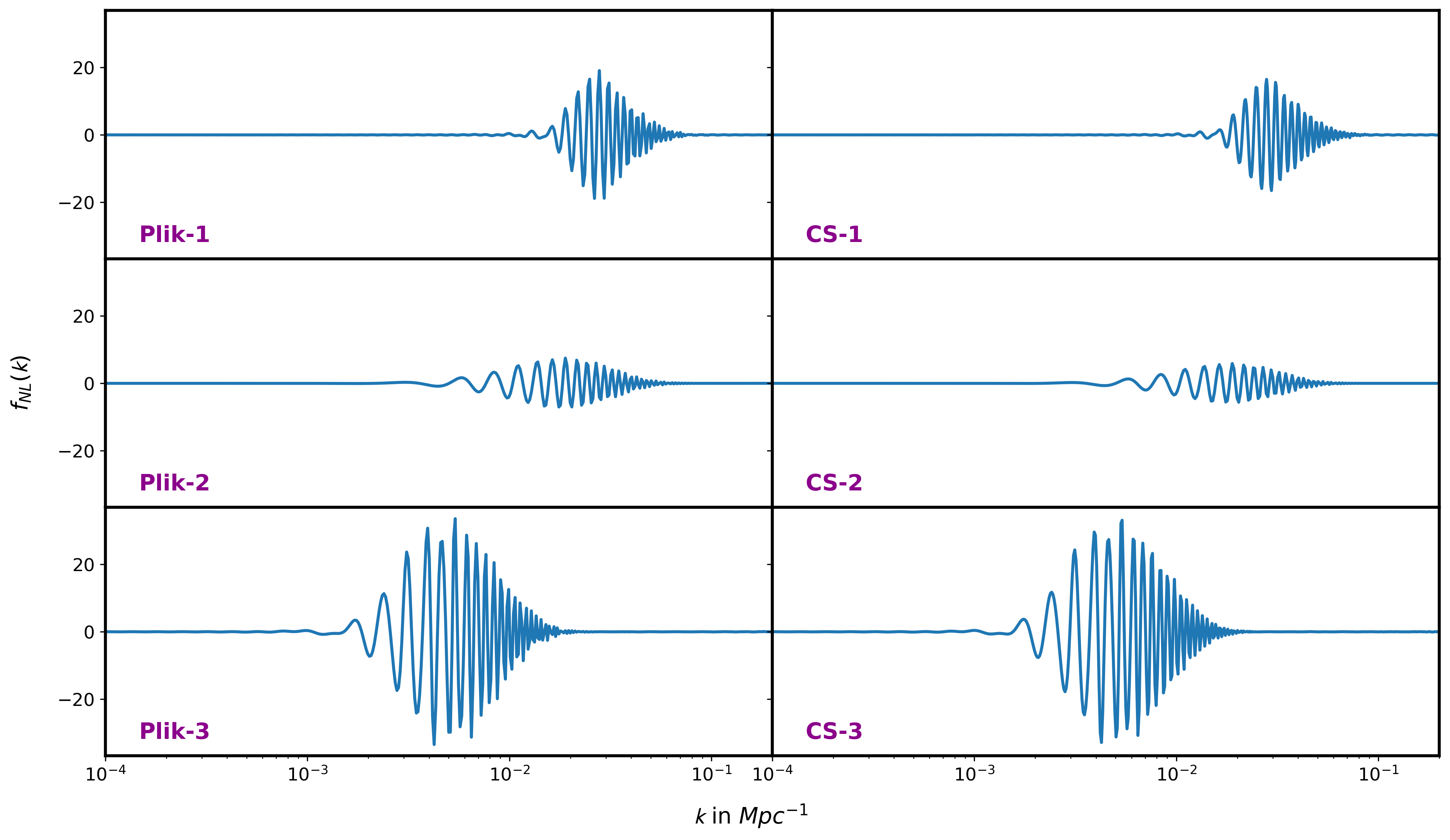}
\caption{Scalar bi-spectrum evaluated in the equilateral limit for the best-fit candidates of the both Plik-bin1 (left side) CamSpec clean  (right side) likelihood. }
\label{equi}
\end{figure*}

\begin{figure*}
\centering
\includegraphics[trim={0 0 0 0},width=0.9\textwidth]{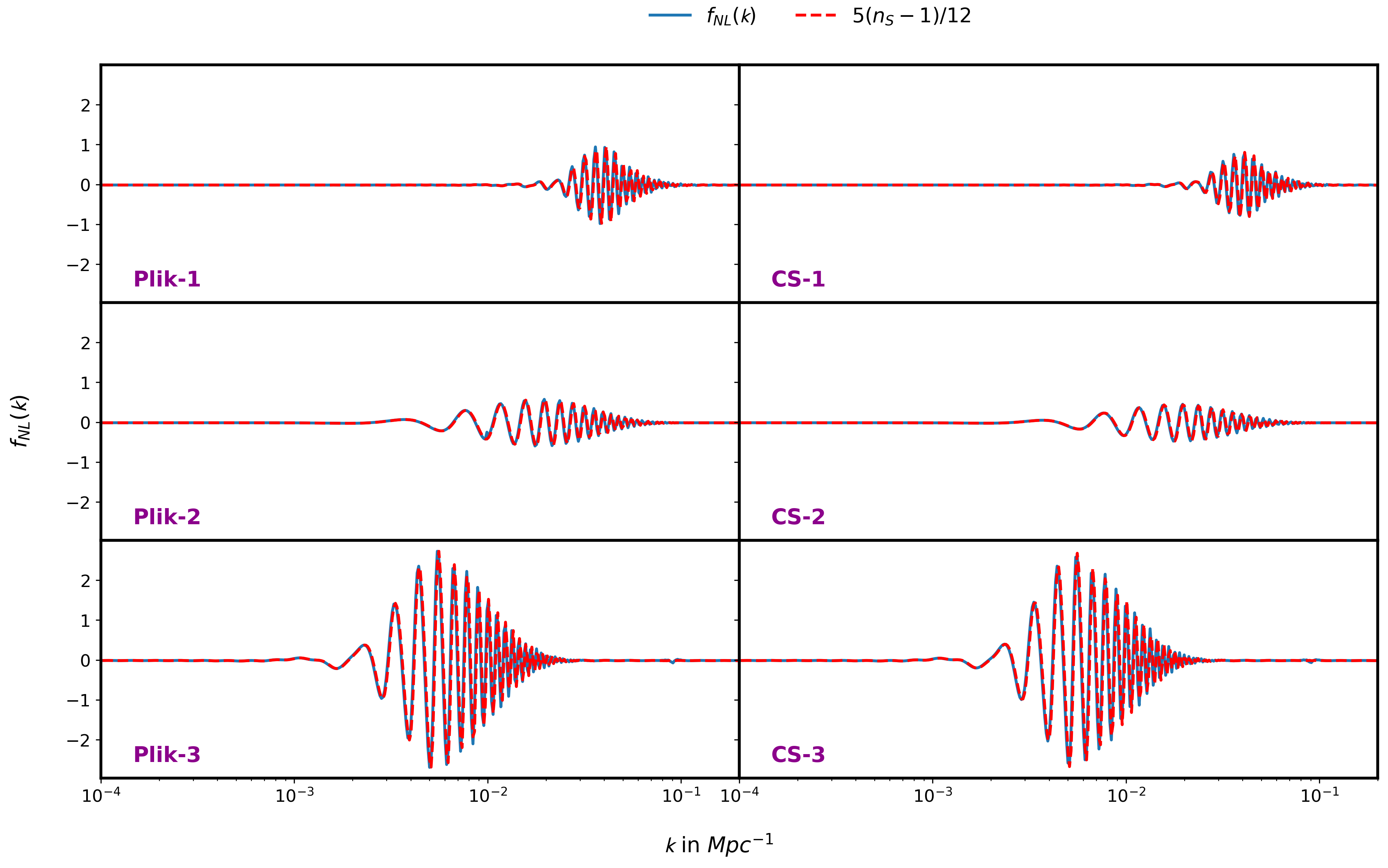}
\caption{Verifying consistency relation (\autoref{cons-eq}) for the best-fit candidates to the Plik-bin1 (left side) and  CamSpec clean (right side) likelihood.}
\label{cons-check}
\end{figure*}

\begin{figure*}
\centering
\includegraphics[trim={0 0 0 0},width=0.9\textwidth]{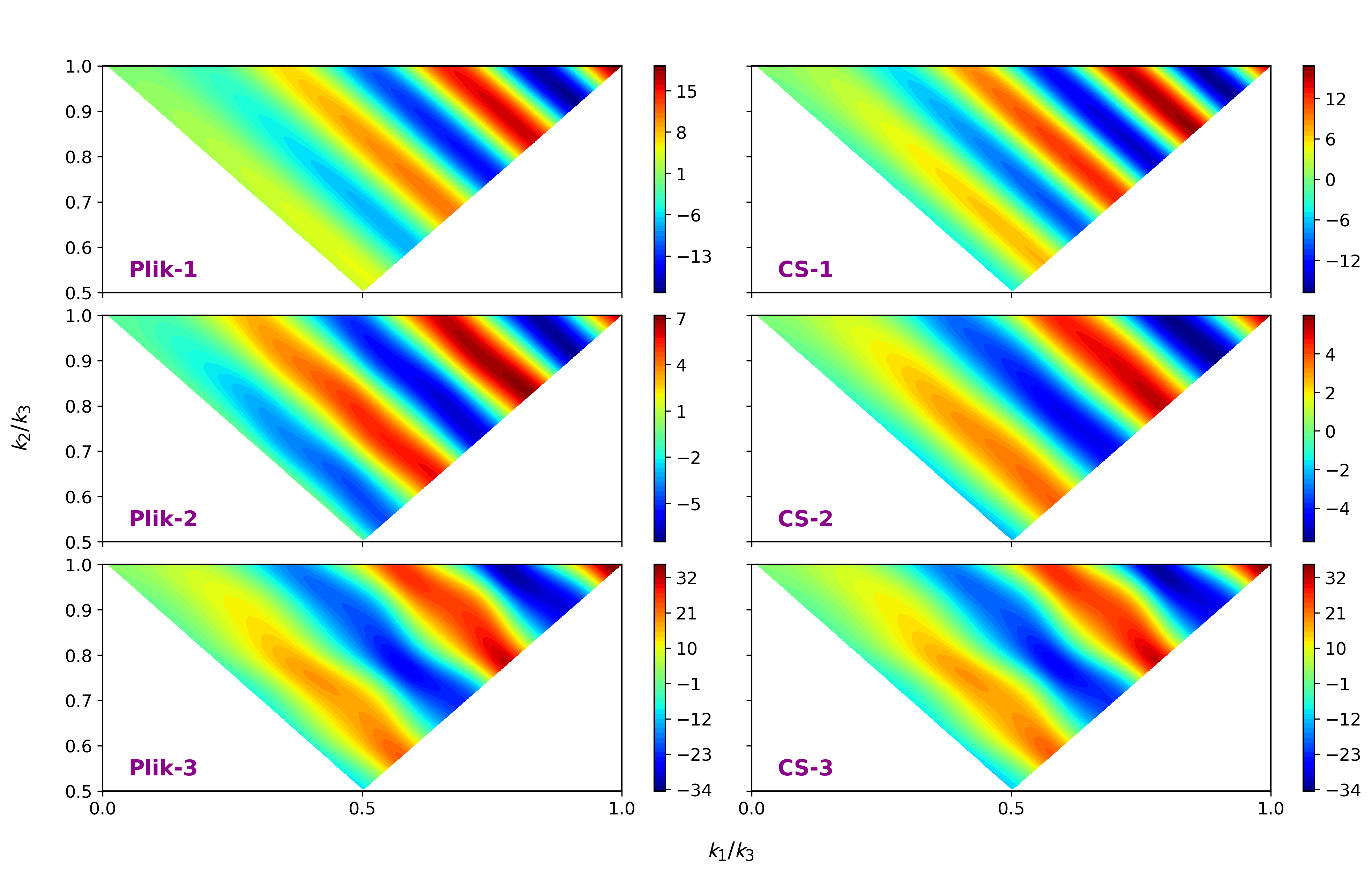}
\caption{2D heat map of $f_{NL}(k_1/k_3,k_2/k_3)$ for the best-fit candidates to the Plik-bin1 (left side) and CamSpec clean (right side) likelihood. $f_{NL}$ values are given in the color bar. Here $k_3$ is the mode which gave the maximum $f_{NL}$ in the equilateral limit for the corresponding candidate.}
\label{scalene}
\end{figure*}

Non-Gaussianity in canonical inflationary models that are completely governed by slow roll dynamics is negligible~\cite{Maldacena_2003,Seery:2005wm}. Deviations from the slow roll nature, on the other hand, produce scale dependent oscillations in the $f_{NL}(k)$~\cite{Chen_2008,Chen:Osc00,Chen:2010xka,Flauger:2010Osc,Chen:2010Osc,Martin:2011sn,Hazra_2013_BINGO,Hazra:2012kq,planck2013_NG,Adshead:2013zfa,Achucarro:2013cva,planck2015_NG,Sreenath_2015,Martin:2014kja,Achucarro:2014msa,Fergusson:2014hya,Meerburg:2015owa,Appleby:2015bpw,Dias:2016rjq,planckcollaboration2019planck}. As a result, features in our candidates produce a significant oscillatory bi-spectrum. In this section, we compute the $f_{NL}$ for each of the six candidates. We are evaluating bi-spectrum in three limits: equilateral ($k_1=k_2=k_3$), squeezed ($k_1\approx k_2 \gg k_3$), and scalene (arbitrary triangular configuration). In all three limits, we use \tw{BINGO} to evaluate the $f_{NL}$. To calculate the $f_{NL}$, we use the same method described in \autoref{PS} and substitute in \autoref{fnl}. \autoref{equi} displays the $f_{NL}$ for all six candidates in the equilateral limits. One can show that in the squeezed limit $f_{NL}$ reduces to
\begin{equation} \label{cons-eq}
    f_{NL} = \frac{5}{12}\left( n_s -1 \right).
\end{equation}
This relation is called the consistency condition \cite{Maldacena_2003, Creminelli}. \autoref{cons-check} verifies the consistency relation for all six candidates. The numerically calculated $f_{NL}$ matches well with the analytical result. In the scalene limit, we obtain the 2D heat map of $f_{NL}(k_1/k_3,k_2/k_3)$ by fixing the value of $k_3$ \cite{Hazra_2013_BINGO}. \autoref{scalene} plots the 2D heat map of $f_{NL}$. Top left corner of 2D map can be identified as the squeezed limit, \ie $k_2=k_3\gg k_1$ and the top right corner is the equilateral limit, \ie $k_1=k_2=k_3$.

One can locate the maximum non-Gaussianity of three point correlation in  equilateral limits (\autoref{equi}). Here, \CS-2 and \pl-2 generates a \fnl  $\sim 6$. This is because of the low frequency oscillations  present in the potential and thereby in the scalar power-spectrum. \CS-3 and \pl-3 generates highest \fnl ~amongst these candidates which is around $\sim 33$. This is due to the presence high amplitude and relatively larger frequency of oscillations present in the potential. Even though \pl-1 and \CS-1 have large frequency oscillations, their amplitude is small compared to other four candidates which  results in a \fnl $\sim 19, 15$ respectively.
\section{Conclusion} \label{conc}
 We have studied the effect of damped oscillations in a  nearly flat inflationary potential and compared the spectra with Planck 2018 data. Our model is able to identify resonant features and sharp feature separately at different scales. We are able see that at  smaller scales data prefers resonant features but at the intermediate scales sharp oscillations give better fit.  We have separately used two high $\ell$ likelihoods, namely Plik-bin1 and CamSpec clean, for the analysis. With the addition of 3 parameters, we are able to get around $10$ and $5$ improvements for Plik-bin1 and CamSpec clean likelihood respectively. The Bayesian analysis didn't give any conclusive evidence for the model. We have studied three candidates each for both Plik-bin1 and CamSpec clean likelihoods. They have provided $10, 8.5, 6$ improvements for the Plik-bin1 and $5, 3.8, 3.7$ for the CamSpec clean. This indicates that the extent of improvement is less in the CamSpec clean compared to the Plik-bin1 for all candidates. While the first candidate has features located at the smaller scales ($10^{-2}-10^{-1}~{\rm Mpc^{-1}}$), the third one has oscillations at large to intermediate scales ($10^{-3}-10^{-2}~{\rm Mpc^{-1}}$). The second candidate has features at the intermediate to small scales. Owing to the feature location for different candidates, different outliers are captured in the $\ell$ space.  Scalar bi-spectrum, \fnl, is evaluated in the squeezed, equilateral limits and also in the scalene configurations. All candidates provide oscillating \fnl with the maximum amplitude reaching up to 33. The first and third candidate have produced relatively higher \fnl amplitude, $\sim 17, 33$, while the second one have generated a maximum \fnl of $\sim 6$ in the equilateral limit. In squeezed limits, the consistency condition is satisfied at all scales for all three candidates of both likelihoods. Along with the constraints on \fnl, we can further narrow down the possible candidates for inflation in future. The feature candidates have overlap with scales that future Large Scale Structure (LSS) probes~\cite{Euclid,LSST,Hazra:2012LSS,Chen:2016LSS,Ballardini:2016LSS,LHuillier:2017lgm,Beutler:2019LSS,Ballardini:2019LSS,Debono20} can explore. Therefore, a joint analysis can help in understanding the significance of these features better.
 
\section*{Acknowledgements}
We would like to thank  Dhiraj Kumar Hazra for the guidance he has given throughout the whole work. We would also like to thank The Institute of Mathematical Sciences (IMSc) for providing all resources needed for the project. All the runs related to this work were done on IMSc's High Performance Computing facility Nandadevi (hpc.imsc.res.in).

\begin{appendices} \label{Appendix}

\section{Theory of inflation} \label{th-inf}
One could engineer an accelerated expansion of the universe very shortly after  big-bang with the help of a single canonical scalar field called inflaton, $\phi(t)$, moving in a potential $V(\phi)$. The action governing such a scalar field is given by
\begin{equation}
    \mc{S}[\phi] = \int \dd^4x\sqrt{-g}\big[X-V(\phi)\big],\label{action_1}
\end{equation}
where $X=-\frac{1}{2}\nabla^\mu\phi\nabla_\mu\phi$. 

Apart from a few simple potential forms, it is not easy to solve the Einstein's equations analytically. In such situations, one can resort to numerical methodologies. For numerical analysis, a better choice of coordinate will be the number of e-folds denoted by $\mc{N}$. It tells number of e-fold times universe expanded in a given cosmic time, \ie $a=a_i\ee^\mc{N}$. The relation between $t$ and $\mc{N}$ is given by
\begin{equation}
    d\mc{N} = H dt,
\end{equation}
where $H$ is the Hubble's parameter in $t$. Evaluating the Einstein's equation corresponding to the FLRW universe (also called Friedmann equations) and the equation of motion of the field, we get
\begin{subequations} \label{fried-eqn_1}
\begin{eqnarray}
   H^2 = &&\frac{V(\phi)}{3-\phi_\mc{N}^2/2}, \\
   \frac{H_\mc{N}}{H} = &&-\frac{\phi_\mc{N}^2}{2},
   \end{eqnarray}
\end{subequations}
\begin{equation}
    \phi_{\mc{NN}} + \phi_\mc{N}\Big(3+\frac{H_\mc{N}}{H}\Big) + \frac{V_\phi}{H^2} = 0, \label{eom_1}
\end{equation}
where subscript $\phi$ denotes the differentiation with respect to $\phi$.

With the help of background equations and linear perturbation theory (\autoref{scalar PS}), we get the governing equation for the curvature perturbation modes as,
\begin{equation}
    \cR_k'' + 2\bigg(\frac{z'}{z}\bigg)\cR_k' + k^2\cR_k = 0,
\end{equation}
where  $z=a\phi'/\mc{H}$. $\mc{H}$ is the Hubble parameter in $\eta$ coordinate. Substituting $\mc{V}=\cR z$ \cite{Sasaki:1986hm,Mukhanov:1987pv,MUKHANOV1992203,Stewart_1993} one can obtain the Mukhanov-Sasaki equation.
\begin{equation}
    \mc{V}_k'' + \bigg[k^2-\Big(\frac{z''}{z}\Big)\bigg]\mc{V}_k = 0.
\end{equation}
Since we will be working in $\mc{N}$ coordinate, these equations can be rewritten as,
\begin{subequations} 
\centering
    \begin{eqnarray}
    {R_k}_{\mc{NN}}+\Big( 1 + \frac{H_\mc{N}}{H} + 2\frac{z_\mc{N}}{z}\Big){R_k}_\mc{N} \\ \nonumber  + \frac{k^2}{a^2H^2}R_k = 0 \label{MSE-N1_1},\\
    {\mc{V}_k}_{\mc{NN}} +\Big( 1 + \frac{H_\mc{N}}{H}\Big){\mc{V}_k}_\mc{N} + \bigg[\frac{k^2}{a^2H^2} \\ \nonumber -\frac{z_{\mc{NN}}}{z}-\frac{z_\mc{N}}{z}\Big(1+\frac{H_\mc{N}}{H}\Big)\bigg]\mc{V}_k = 0 .\label{MSE-N2_1}
    \end{eqnarray}
\end{subequations}
Quantizing the above equation, one can calculate the power-spectrum and is given by the expression,
\begin{equation} \label{PS-eqn_1}
   \mc{P}_s(k) = \frac{k^3}{2\pi^2}\abs{\cR_k}^2. 
\end{equation}
Finally one can calculate the angular power-spectrum from the primordial power-spectrum using the following relation \cite{Hu_2004},
\begin{equation}
    \cC_l^{XY} = \frac{2}{\pi}\int k^2\dd k \cP_s(k)\mc{T}^{Xl}(k)\mc{T}^{Yl}(k),
\end{equation}
where $\mc{T}$ is the transfer function which is calculated using the Boltzmann equations.

The fluctuations in $\mc{R}$ might not be always Gaussian in nature, there might be some considerable amount of non-Gaussianities present in it. With the help of 3-point correlation of curvature perturbation, one can obtain the measure of non-Gaussianity, \fnl, to be (\autoref{scalar BS}) 
\begin{align} \label{fnl_1}
f_{NL}(\textbf{k}_1,\textbf{k}_2,\textbf{k}_3) = &-\frac{10}{3}(2\pi)^{-4} (k_1k_2k_3)^3G(\textbf{k}_1,\textbf{k}_2,\textbf{k}_3)\nonumber \\
&\times[k_1^3\mathcal{P}_S(k_2)\mathcal{P}_S(k_3) + \mbox{two permutations}]^{-1}.
\end{align}
Using Maldacena formalism \cite{Maldacena_2003}, $G(k_1,k_2,k_3)$ can be expressed as
\begin{equation} 
\begin{aligned}
G(\textbf{k}_1,\textbf{k}_2,\textbf{k}_3) 
               &= \sum^6_{C=1}[\mathcal{R}_{k_1}(\mc{N}_e)\mathcal{R}_{k_2}(\mc{N}_e)\mathcal{R}_{k_3}(\mc{N}_e)]\mathcal{G}_C(\textbf{k}_1,\textbf{k}_2,\textbf{k}_3)\\
               +&[\mathcal{R}_{k_1}^*(\mc{N}_e)\mathcal{R}_{k_2}^*(\mc{N}_e)\mathcal{R}_{k_3}^*(\mc{N}_e)]\mathcal{G}_C^*(\textbf{k}_1,\textbf{k}_2,\textbf{k}_3)]\\
               +&G_7(\textbf{k}_1,\textbf{k}_2,\textbf{k}_3).
\end{aligned}
\end{equation}

\section{Scalar power-spectrum} \label{scalar PS}

 Like in any other perturbation theory, we  decompose our variable of interest into background quantity and a variation then equate them separately. Therefore, the Einstein's equation at the first order of variation can be written as
\begin{equation} \label{delta_eins_eq}
\delta G_{\mu\nu} = 8\pi G\delta T_{\mu\nu}.
\end{equation}
This is, in fact, a set of linear differential equation that governs the dynamics of perturbation in metric ($\delta g_{\mu\nu}$). In the FLRW background, based on the response to a local rotation of spatial coordinates on constant time hyper-surface, one can decompose the perturbations into scalar, vector, and tensor transformations. Out of these, only scalar power-spectrum has re-entered the Hubble horizon during the recombination epoch and is responsible for the anisotropies present in the universe. Since our goal is to account for anisotropies in the CMB data, we will be working with scalar perturbations only.
 On taking into account the scalar perturbations to  background metric, the FLRW line element can be written as~\cite{MUKHANOV1992203,Sasaki:1986hm}
\begin{align}
{\rm d} s^2
= -\left(1+2\, A\right)\,{\rm d} t ^2 
+ 2\, a(t)\, (\partial_{i} {D})\; {\rm d} t\; {\rm d} x^i\, \\ \nonumber +a^{2}(t)\; \left[(1-2\, B)\; \delta _{ij}
+ 2\, \left(\partial_{i}\, \partial_{j}{E} \right)\right]\,
{\rm d} x^i\, {\rm d} x^j. \label{eq:f-le-sp_1}
\end{align}

There are two approaches to study the evolution of perturbations. One is to construct gauge invariant quantities and study using them while the other is to choose a specific gauge and work throughout in it. We will be working in a specific gauge, longitudinal gauge. Because of the coordinate freedom we have, we set $D=E=0$. Therefore metric in the new gauge can be written as,   
\begin{equation}
{\rm d} s^2
= -\left(1+2\, A\right)\,{\rm d} t ^2 
+a^{2}(t)\; \left[(1-2\, B)\; \delta _{ij}
\right]\,
{\rm d} x^i\, {\rm d} x^j. \label{eq:f-le-sp_2}
\end{equation}
Using the above metric, the independent components of the Einstein's tensor can be found to be \cite{Bassett}
\begin{subequations} \label{Eins-tensor}
\begin{eqnarray}
\delta G^0_0 &&= \;6H(\Dot{B}+HA) -\frac{2}{a^2}\nabla^2B,\\
\delta G^0_i &&= \;-2\nabla_i(\Dot{B} + HA), \\
\delta G^i_j &&= \;2\big[\Ddot{B} + H(3\Dot{B}+\Dot{A}) + (2\Dot{H} + 3H^2) A \\\nonumber
&&\,\,   \;+\frac{1}{a^2}\nabla^2(A-B)\big]\delta^i_j + \frac{1}{a^2}\nabla^i\nabla_j(B-A).
\end{eqnarray}
\end{subequations}

If we consider $\delta\phi$ to be the perturbation in scalar field, then the perturbed  energy momentum tensor, up to linear order, can be written as 
\begin{subequations} \label{tmunu}
\begin{eqnarray}
\delta T^0_0 = &&-(\Dot{\phi}\Dot{\delta\phi} -\Dot{\phi}^2A + V_\phi\delta\phi), \\
\delta T^0_i = &&-\nabla_i(\Dot{\phi}\delta\phi), \\
\delta T^i_j = &&(\Dot{\phi}\Dot{\delta\phi} -\Dot{\phi}^2A - V_\phi\delta\phi)\delta^i_j.
\end{eqnarray}
\end{subequations}
In the absence of anisotropic stresses $\delta T^i_j=0$ for $i\neq j$,  \ie from \autoref{delta_eins_eq} we get $A=B$. Therefore, we can rewrite the Einstein equation from \autoref{Eins-tensor} and \autoref{tmunu} in the conformal coordinates as follows \cite{MUKHANOV1992203},
\begin{equation} \label{bardeen-eqn}
    A'' + 6\mc{H}A'-\nabla^2A + (2\mc{H}'+4\mc{H}^2)A =0,
\end{equation}
where $\mc{H}$ is the Hubble parameter in conformal coordinates. Since we are analysing a canonical model,  speed of perturbation is equal to 1. Also we assume that perturbations are purely adiabatic in nature \cite{Gordon_2000}, \ie $\delta T^0_0 = \delta T^i_j$. Now defining curvature perturbation as \cite{MUKHANOV1992203} 
\begin{equation}\label{curv-per}
    \cR = A + \frac{2\rho}{3\mc{H}}\frac{A' + \mc{H}A}{\rho + P},
\end{equation}
and with the help of background equations, \autoref{fried-eqn_1}, we get  $\cR_k'$ in the Fourier space to be 
\begin{equation}
    \cR_k' = -\frac{\mc{H}}{\mc{H}^2-\mc{H}'}k^2A_k.
\end{equation}

Differentiating the above equation with respect to the conformal time and with the help \autoref{bardeen-eqn} and \autoref{curv-per}, we get the curvature perturbation equation given by  
\begin{equation}
    \cR_k'' + 2\bigg(\frac{z'}{z}\bigg)\cR_k' + k^2\cR_k = 0,
\end{equation}
where  $z=a\phi'/\mc{H}$. Introducing a new variable $\mc{V}=\cR z$ \cite{Sasaki:1986hm,Mukhanov:1987pv,MUKHANOV1992203,Stewart_1993} we have,
\begin{equation}
    \mc{V}_k'' + \bigg[k^2-\Big(\frac{z''}{z}\Big)\bigg]\mc{V}_k = 0.
\end{equation}
This equation is called the Mukhanov-Sasaki equation.
Quantizing the above equation, one can calculate the power-spectrum and is given by the expression,
\begin{equation} \label{PS-eqn}
   \mc{P}_s(k) = \frac{k^3}{2\pi^2}|{\cR_k}|^2. 
\end{equation}

\section{Scalar bi-spectrum} \label{scalar BS}
 To obtain the non-Gaussianities present in $\mc{R}$ one need to calculate the scalar bi-spectrum, $\mathcal{B}_S(\textbf{k}_1,\textbf{k}_2,\textbf{k}_3)$ \cite{Maldacena_2003}, at end of inflation($\mc{N}_e$)  which is defined in terms of three point correlation function of $\mc{R}$ as \cite{Seery_2005,Chen_2005,Chen_2008, Martin_2012}
\begin{align} \label{bispec}
\langle\hat{\mathcal{R}}_{\textbf{k}_1}\hat{\mathcal{R}}_{\textbf{k}_2}\hat{\mathcal{R}}_{\textbf{k}_3}\rangle = (2\pi)^3\mathcal{B}_S(\textbf{k}_1,\textbf{k}_2,\textbf{k}_3)\delta^{(3)}(\textbf{k}_1+\textbf{k}_2+\textbf{k}_3). 
\end{align}
One can adopt the following ansatz to study non-Gaussinities \cite{Komatsu_2001},
\begin{align}
\mathcal{R}(\mc{N},\textbf{k}) = \mathcal{R}_G(\mc{N},\textbf{k})-\frac{3f_{NL}}{5}\left[\mathcal{R}_G^2(\mc{N},\textbf{k})-\langle\mathcal{R}_G^2(\mc{N},\textbf{k})\rangle\right],
\end{align}
where $\mathcal{R}_G$ is the the Gaussian part and $f_{NL}$ is the non-Gaussianity parameter. With the help of Wick's theorem, the 3-point correlation of the curvature perturbation can be evaluated as 
\begin{equation}
\begin{aligned} \label{3point_corr}
\langle\hat{\mathcal{R}}_{\textbf{k}_1}\hat{\mathcal{R}}_{\textbf{k}_2}\hat{\mathcal{R}}_{\textbf{k}_3}\rangle &=-\frac{3f_{NL}}{10}(2\pi)^4(2\pi)^{-3/2} \frac{1}{k_1^3k_2^3k_3^3}\\\times&\delta^{(3)}(\textbf{k}_1+\textbf{k}_2+\textbf{k}_3)\\
\times &[k_1^3\mathcal{P}_S(k_2)\mathcal{P}_S(k_3) + \mbox{two permutations}].
\end{aligned}
\end{equation}
Using the ansatz $\mathcal{B}_S(\textbf{k}_1,\textbf{k}_2,\textbf{k}_3) = (2\pi)^{-9/2} G(\textbf{k}_1,\textbf{k}_2,\textbf{k}_3)$,  \autoref{bispec} and \autoref{3point_corr}, $f_{NL}$ can be determined to be the following
\begin{align} \label{fnl}
f_{NL}(\textbf{k}_1,\textbf{k}_2,\textbf{k}_3)&= -\frac{10}{3}(2\pi)^{-4} (k_1k_2k_3)^3G(\textbf{k}_1,\textbf{k}_2,\textbf{k}_3)\nonumber \\
\times[k_1^3&\mathcal{P}_S(k_2)\mathcal{P}_S(k_3) + \mbox{two permutations}]^{-1}.
\end{align}
Using Maldacena formalism \cite{Maldacena_2003}, $G(k_1,k_2,k_3)$ can be expressed as
\begin{equation*} 
\begin{aligned}
G(\textbf{k}_1,\textbf{k}_2,\textbf{k}_3) 
               &= \sum^6_{C=1}\Big[[\mathcal{R}_{k_1}\mathcal{R}_{k_2}\mathcal{R}_{k_3}]\mathcal{G}_C(\textbf{k}_1,\textbf{k}_2,\textbf{k}_3)\\
               &+[\mathcal{R}_{k_1}^*\mathcal{R}_{k_2}^*\mathcal{R}_{k_3}^*]\mathcal{G}_C^*(\textbf{k}_1,\textbf{k}_2,\textbf{k}_3)\Big]\\
                               &+G_7(\textbf{k}_1,\textbf{k}_2,\textbf{k}_3),
\end{aligned}
\end{equation*}
where each individual term is given by the following:

\begin{flalign*}
\mathcal{G}_1(\textbf{k}_1,\textbf{k}_2,\textbf{k}_3) &= 2i\int^{\mc{N}_e}_{\mc{N}_i} \mbox{d}\mc{N} a^3\epsilon_1^2H\mathcal{R}_{k_1}^*(\mathcal{R}_{k_2}^*)_\mc{N}(\mathcal{R}_{k_3}^{*})_\mc{N}\\ \nonumber &+ \mbox{two permutations}),\\
\mathcal{G}_2(\textbf{k}_1,\textbf{k}_2,\textbf{k}_3) &= -2i(\textbf{k}_1.\textbf{k}_2+\mbox{two permutations})\\ \nonumber
&\times\int_{\mc{N}_i}^{\mc{N}_e} \mbox{d}\mc{N} \frac{a \epsilon_1^2}{H} \mathcal{R}_{k_1}^*\mathcal{R}_{k_2}^*\mathcal{R}_{k_3}^*,\\
\mathcal{G}_3(\textbf{k}_1,\textbf{k}_2,\textbf{k}_3) &= -2i\int_{\mc{N}_i}^{\mc{N}_e} \mbox{d}\mc{N} a^3H \epsilon_1^2\Bigg[\frac{\textbf{k}_1.\textbf{k}_2}{k_2^2}\times\\ \nonumber
&\mathcal{R}_{k_1}^*(\mathcal{R}_{k_2}^{*})_\mc{N}(\mathcal{R}_{k_3}^{*})_\mc{N} + \mbox{five permutations}\Bigg],\\
\mathcal{G}_4(\textbf{k}_1,\textbf{k}_2,\textbf{k}_3) &= i \int_{\mc{N}_i}^{\mc{N}_e} \mbox{d}\mc{N} a^3 \epsilon_1 (\epsilon_2)_\mc{N} H \Big[\mathcal{R}_{k_1}^*\mathcal{R}_{k_2}^*(\mathcal{R}_{k_3}^{*})_\mc{N} \\ \nonumber
&+ \mbox{two permutations}\Big],\\
\mathcal{G}_5(\textbf{k}_1,\textbf{k}_2,\textbf{k}_3) &= \frac{i}{2} \int_{\mc{N}_i}^{\mc{N}_e} \mbox{d}\mc{N} a^3 \epsilon_1^3H \Bigg[\left(\frac{\textbf{k}_1.\textbf{k}_2}{k_2^2}\right)\times \\ \nonumber &\mathcal{R}_{k_1}^*(\mathcal{R}_{k_2}^{*})_\mc{N}(\mathcal{R}_{k_3}^{*})_\mc{N}+ \mbox{five permutations}\Bigg]\\
\mathcal{G}_6(\textbf{k}_1,\textbf{k}_2,\textbf{k}_3) &= \frac{i}{2} \int_{\mc{N}_i}^{\mc{N}_e} \mbox{d}\mc{N} a^3\epsilon_1^3H \Bigg[\left(\frac{k_1^2(\textbf{k}_2.\textbf{k}_3)}{k_2^2k_3^2}\right)\times\\&\mathcal{R}_{k_1}^*(\mathcal{R}_{k_2}^{*})_\mc{N}(\mathcal{R}_{k_3}^{*})_\mc{N}+\mbox{two permutations}\Bigg],\\
G_7(\textbf{k}_1,\textbf{k}_2,\textbf{k}_3) &= \frac{\epsilon_2(\mc{N}_e)}{2}\Big[
]|\mathcal{R}_{k_2}(\mc{N}_e)|^2|\mathcal{R}_{k_3}(\mc{N}_e)|^2 \\ &+ \mbox{two permutations}\Big].
\end{flalign*}
where $\epsilon_2$ is the second slow roll parameter that is defined with respect to the first as follows: $\epsilon_2 = \mbox{d ln} \epsilon_1/\mbox{d}\mc{N}$ \cite{Liddle_1994}.

\end{appendices}

\bibliographystyle{JHEP}
\bibliography{main}
\end{document}